\documentclass[11pt, onecolumn, draftcls ]{IEEEtran}
\usepackage{cite}
\usepackage{algorithm2e}
\usepackage{amssymb}
\usepackage{enumerate}
\usepackage{graphicx}
\usepackage{amssymb}
\usepackage{amsbsy}
\usepackage{float}
\usepackage{balance}
\usepackage{subfigure}
\usepackage{url}
\usepackage{bm}
\newfloat{longequation}{b}{ext}
\newfloat{longequation}{t}{ext}

\usepackage[cmex10]{amsmath}
\graphicspath{{./figs/}}
\usepackage{amssymb}
\usepackage{graphicx}
\usepackage{amssymb}
\usepackage{amsbsy}
\usepackage{float}
\usepackage{balance}
\usepackage{subfigure}
\usepackage{url}
\usepackage{bm}
\newfloat{longequation}{b}{ext}
\newfloat{longequation}{t}{ext}
\usepackage{array}
\usepackage{varioref}
\usepackage[normalem]{ulem}
\graphicspath{{./figs/}}
\usepackage{cite}
\usepackage{color}
\usepackage{empheq}
\begin{document}

\title{Multicast Multigroup Beamforming under Per-antenna Power Constraints}
\author{\IEEEauthorblockN{ Dimitrios Christopoulos\IEEEauthorrefmark{1},   Symeon Chatzinotas\IEEEauthorrefmark{1}
 and Bj\"{o}rn Ottersten\IEEEauthorrefmark{1}\\
 \thanks{This work was   partially supported by the National Research Fund, Luxembourg under the project  ``$\mathrm{ CO^{2}SAT}$: Cooperative \& Cognitive Architectures for Satellite Networks'.}
 }

\IEEEauthorblockA{\IEEEauthorrefmark{1}SnT - securityandtrust.lu,  University of Luxembourg
\\email: \textbraceleft dimitrios.christopoulos, symeon.chatzinotas, bjorn.ottersten\textbraceright@uni.lu}

%
}
\maketitle


\begin{abstract}
Linear precoding exploits the spatial degrees of freedom offered by multi-antenna transmitters to serve multiple  users over the same frequency resources. The present work focuses on  simultaneously serving multiple groups of users, each with its own channel, by transmitting  a stream of  common symbols  to each  group. This scenario is known as physical layer multicasting to multiple co-channel groups. Extending the current state of the art in multigroup multicasting, the practical constraint of a maximum permitted  power level  radiated by each antenna is tackled herein. The considered per antenna power constrained system is optimized in a  maximum fairness sense. In other words,  the optimization aims at favoring the worst user by  maximizing the minimum rate. This Max-Min Fair criterion is imperative in multicast systems, where   the performance of all the receivers listening to the same  multicast is dictated by the worst rate in the group. An analytic framework to tackle  the Max-Min Fair multigroup multicasting scenario under per antenna power constraints is therefore derived. Numerical results display the accuracy of the proposed solution and provide insights to the performance of a per antenna power constrained system.       \end{abstract}
\section{Introduction}
Multiuser multiple antenna transmitters are the way forward towards achieving the high throughput demands of next generation systems. Advanced transmit signal processing techniques are employed to optimize the performance of the multi-antenna transmitter without compromising the complexity of the receivers. A fundamental requisite for the application of these techniques, namely linear transmit beamforming (also reffered to as precoding) is the perfect knowledge of the channel state at the transmitter. 
Subsequently, the exploitation of the spatial degrees of freedom  offered by the  antenna array mitigates interferences thus allowing co-channel beams
 to be made adjacent. In this fashion, a Spatial Division Multiple Access (SDMA) scheme is realized.

Physical layer multicasting has the potential to efficiently address the nature of traffic demand in future systems and  has become part of standards such as LTE. When multiple multicast co-channel groups are considered towards trading off between the unicasting and broadcasting system functionalities,   then the design can be optimized in a  manner  described hereafter.

An important application of physical layer multigroup multicasting can be found when the goal is to  optimize full frequency reuse multibeam transmitters without changing the framing structure of communication standards. For instance, specific physical layer designs are optimized to cope with noise limited channels with long propagation delay. What is more,   the framing of multiple users per transmission is  emanated  to guarantee scheduling efficiency when long forward error correction codes are employed.  Consequently,  precoding techniques in such systems cannot be based on the conventional user-by-user design and multigroup multicasting needs to be considered.

\section{Related Work}
In the multigroup multicasting literature, two fundamental optimization problems have been considered until now; the sum power minimization under specific quality of service ($\mathrm{QoS}$) constraints and the maximization of the minimum $\mathrm{SNIR}$ (\textit{Max-Min Fair}).  In the following, a brief literature review is provided
\subsection{Sum Power Minimization under $\mathrm{QoS}$ constraints}

 In multi-antenna systems, the optimal downlink transmission strategy, in the sense of minimum total transmit power that guarantees specific $\mathrm{QoS}$ constraints at each user,    was given in \cite{Bengtsson2001,bengtsson1999}.  Therein, the powerful tool of Semi-Definite Relaxation ($\mathrm{SDR}$) reduced the non-convex quadratically constrained quadratic  program ($\mathrm{QCQP}$)  into a relaxed convex  problem by changing the optimization variables and disregarding the unit-rank constraints over the new variable. One fundamental  assumption of this work has been the  independence of the data addressed to the multiple users. Under this assumption,   the $\mathrm{SDR}$   guarantees an optimal solution to the $\mathrm{QCQP}$ problem\cite{Bengtsson2001}.

 Based on the principles of $\mathrm{SDR}$, the $\mathrm{QoS}$ problem in a physical layer multicasting scenario\footnote{The multi-antenna multicast problem was originally proposed in \cite{Lopez} where the maximization of the average $\mathrm{SNIR}$ over all users was the goal,  without however guaranteeing some $\mathrm{QoS}$ at the user side.}was tackled by \textit{Sidiropoulos et al.} in\cite{Sidiropoulos2006}. The inherent difference  between the prior scenario of transmitting  independent information to each user, is that in the physical layer multicasting problem, a set of  \textit{common} symbols is addressed to all the users. The general multicasting problem was  proven NP-hard \cite{Sidiropoulos2006}. This result suggested that if the polynomial in complexity $\mathrm{SDR}$ method could provide the globally optimal solution, then it would be possible to solve a whole class of computationally challenging problems in polynomial time.
Consequently and in order to derive a low complexity and accurate approximate solution
to the multicast problem, $\mathrm{SDR}$ was combined with Gaussian randomization  \cite{Luo2010}. In more detail, a candidate solution to the original problem is generated as a random Gaussian variable with statistics given by the solution of the relaxed problem. After generating a finite number of random Gaussian instances, depending on the desired accuracy of the solution, the instance that yields the best objective value of the original problem is chosen. A key step in this process remains the simple power rescaling of the random Gaussian solutions which guarantees the feasibility of the original problem without affecting the  beamforming directions.

A unified framework for physical layer multicasting to multiple interfering groups was given in \cite{Karipidis2005CAMSAP,Karipidis2008}. In this case, independent sets of common data are transmitted to  multiple interfering groups of users by the multiple antennas. Subsequently, common information is now restricted to users that belong to the same group while independent information is sent to different groups.  In the extreme cases of  either one user per group  or a single group that includes all the users, the  model reduces to the independent data or the multicast  scenarios respectively. This general problem was also shown to be  NP-hard to solve, since it includes an NP-hard problem as a special case\cite{Karipidis2008}. Despite the existence of  an optimal solution for the single user per group case  of the multicast multigroup scenario via $\mathrm{SDR}$, a general solution to the multicast multigroup beamforming optimization problem is more complicated. The combination of $\mathrm{SDR}$ and Gaussian randomization is also not straightforward. The difficulty lies in the coupling  between the multicast groups. More specifically,  the issue of intergroup interference arises due to the independent data being transmitted to different groups over the same channels. Hence, in contrast to the multicast case, after obtaining candidate solutions from the Gaussian randomization process  the feasibility of the initial $\mathrm{QCQP}$ cannot be guaranteed by a simple rescaling of the randomized precoding vectors. Since the the  co-channel groups are coupled by interferences, rescaling the power of one precoder invalidates the solution.
To the end of solving the elaborate multicast multigroup problem, \textit{Karipidis et al} \cite{Karipidis2008} proposed an additional step following the Gaussian randomization. This step consists of a new optimization problem, namely the \textit{multigroup multicast  power control} ($\mathrm{MMPC}$),  that converts the candidate solutions into feasible solutions of the original problem\cite{Karipidis2008,Karipidis2005CAMSAP}. This   power control  is a  linear program that guarantees the feasibility of the original problem.

For the sake of completeness, in parallel to \cite{Karipidis2005CAMSAP}, the independent work   presented in \cite{Gao2005} is pointed out, where  the multicast multigroup problem is also tackled. Despite the use of the $\mathrm{SDR}$ method combined with Gaussian randomization to solve the $\mathrm{QoS}$ problem, \textit{Gao and Schubert}\cite{Gao2005} rely on dirty paper coding $\mathrm{DPC}$ to successively precancel intergroup interferences,
thus simplifying the power allocation procedure. Nevertheless, the design of such a  transmitter suffers from the non-linear implementation complexity.
\subsection{Max-Min Fair beamforming under Sum-Power constraints}
 The maximization of the minimum signal to noise plus interference ratio $\mathrm{(SNIR)}$ received by any of the available users in the coverage area, subject to a sum  power constraint $(\mathrm{SPC})$ at the transmitter is a problem closely related to the $\mathrm{QoS}$ problem. The goal of this problem is to maximize the fairness of the system by boosting the $\mathrm{SNIR}$ of the worst user. However, this is accomplished at the expense of the total system sum-rate, otherwise achievable by favouring the well conditioned users.  As proven in \cite{Sidiropoulos2006},  a $\mathrm{QoS}$ problem  with equal  $\mathrm{SNIR}$ constraints is equivalent to the Max-Min-Fair problem up to scaling. Thus, after establishing that the latter is also NP-hard to solve, a customized low complexity algorithm for an approximate solution to the fairness problem was derived \cite{Sidiropoulos2006}. In the same direction, the maximum fairness problem was also formulated, proven NP-hard and solved in a  multicast multigroup scenario in \cite{Karipidis2008}. In more detail, the equivalence the  $\mathrm{QoS}$ and the Max-Min-Fair problems is used to solve the latter. A bisection search method over the relaxed power minimization problem provides candidate solutions to the initial fairness problem. However, feasibility of the original problem with these solutions is not guaranteed and an additional power allocation needs to  come in play. The complication in the fairness scenario  is that the \textit{multigroup multicast  power control} program does not admit a linear program reformulation and thus its solution is not trivial\footnote{It can however be reformulated as a geometric problem ($\mathrm{GP}$) and solved efficiently with interior point methods \cite{Karipidis2008}.}. To overcome this, a bisection search algorithm is again performed, this time over the multigroup multicast power control of the power minimization problem.

In optimization terminology, the fact that the $\mathrm{SDR}$  can provide a global optimum  for the original unicast problem shows that strong duality holds. In this light, the authors of \cite{Silva2009}, tackled the multicast multigroup problem under $\mathrm{SPC}$, based on the framework of uplink-downlink duality \cite{Schubert2004}. Since in the multigroup multicast case one precoder needs to apply for multiple users and thus the principles of duality cannot be straightforwardly applied,    approximations where employed. Subsequently,  low complexity \textit{multicast-aware} solutions where proposed in  \cite{Silva2009}. These solutions where shown to tradeoff the low complexity with the inferior performance in terms of $\mathrm{BER}$ (i.e. minimum $\mathrm{SNIR}$) compared to \cite{Karipidis2008}, under numerically exhibited convergence.

Under a different system model,  the solution of the Max-Min Fair problem  is given in \cite{Xiang2013}, for coordinated  multicast multi-cell systems. In this scenario, a sum power constraint is no longer applicable hence the the $\mathrm{QoS}$ problem is no longer related to the max-min-fair problem with per base-station ($\mathrm{BS}$) constraints. Nevertheless, in each $\mathrm{BS}$ a single group of receivers is assumed. Hence, a power constraint over each precoder is imposed. A solution of the new optimization problem was given following the well established framework of bisection by a modification of the related $\mathrm{QoS}$ problem.
%

 Despite the extensive literature on the topic of multigroup multicasting,  no existing work has tackled the multigroup multicast problem under per antenna power constraints $\mathrm{(PAC)}$. The per-antenna constraints are commonly introduced by the practical limitations of transmitters. When power sharing of the per antenna dedicated $\mathrm{RF}$ chains is not possible, then the $\mathrm{SPC}$ is no longer applicable.
For the sake of clarity, the difference between coordinated and cooperative   multicell networks is pointed out.  In cooperative multicast multicell  systems, all $\mathrm{BS}$s jointly transmit to multicast groups. This varies from the coordinated case of  \cite{Xiang2013} where each $\mathrm{BS}$ transmits to a single multicast group.

The remaining of the paper is structured as follows. The multigroup multicast system model is presented in Sec. \ref{sec: System Model}. The Max-Min Fair under $\mathrm{PAC}$ problem formulation is given in Sec. \ref{sec: problem} along with a detailed solution. Numerical results are provided in Sec. \ref{sec: simulations} while the conclusions are drawn in Sec. \ref{sec: conclusions}\footnote{\textit{Notation}: Throughout the paper, bold face lower case characters denote column vectors and upper case denote matrices. The operators \(\left(\cdot\right)^\dag\), $|\cdot|$ and \(||\cdot||_2, \) denote   the conjugate transpose, the absolute value and the Frobenius norm operations, respectively, while $[\cdot]_{ij}  $  denotes the $i, j$-th element of a matrix.  The principal eigenvalue of a matrix $\mathbf X$ are denoted as $\lambda_{max}(\mathbf X)$.}.


\section{System Model}\label{sec: System Model}
 The focus is on a multi-user ($\mathrm{MU}$) multiple input single output ($\mathrm{MISO}$) Multicast system. Assuming a single transmitter, let   $N_t$ denote the number of transmitting elements  and by $N_{u}$ the  number of users.  The input-output analytical expression  will read as $y_{i}= \mathbf h^{\dag}_{i}\mathbf x+n_{i},$
where \(\mathbf h^{\dag}_{i}\) is a \(1 \times N_{t}\) vector composed of the channel coefficients (i.e. channel gains and phases) between the \(i\)-th user and the  \(N_{t}\) antennas of the transmitter, \(\mathbf x\) is the \(N_{t} \times 1\)  vector of the transmitted signal and  \(n_{i}\) is the independent complex circular symmetric (c.c.s.) independent identically distributed (i.i.d) zero mean  Additive White Gaussian Noise ($\mathrm{AWGN}$)  measured at the \(i\)-th user's receive antenna.


Focusing  in a multigroup multicasting scenario,  let there be a total of $1\leq G \leq N_{u}$ multicast groups with  $\mathcal{I} = \{\mathcal{G}_1, \mathcal{G}_2, \dots  \mathcal{G}_G\}$ the collection of   index sets and $\mathcal{G}_k$ the set of users that belong to the $k$-th multicast group, $k \in \{1\dots G \}. $ Each user belongs to only one group, thus $\mathcal{G}_i\cap\mathcal{G}_j=\O ,  \forall i,j \in \{1\cdots G\}$. Let $\mathbf w_k \in \mathbb{C}^{N_t \times 1}$ denote the precoding weight vector applied to the transmit antennas to beamform towards the $k$-th group. By assuming independent information transmitted to different groups, the symbol streams $\{s_k\}_{k=1}^G$ are mutually uncorrelated and the total power radiated from the antenna array is equal to $P_{tot} = \sum_{k=1}^ G ||\mathbf w_k||^2_2$. The power radiated by each antenna element is also a linear combination of all precoders and reads as
\cite{Yu2007}\begin{align}\label{eq: PAC}
P_n = \left[\sum_{k=1}^G \mathbf w_k \mathbf w_k^\dag \right]_{nn} ,
\end{align}
where $n \in \{1\dots  N_t\}$.
In \eqref{eq: PAC} the fundamental difference between the $\mathrm{SPC}$  of \cite{Karipidis2008} and the proposed $\mathrm{PAC}$ is pointed out. Herein, instead of one,  $N_t$ constraints are imposed, each one involving all the precoders. The main complication of this is that   the $\mathrm{QoS}$ problem  is no longer related to the Max-Min problem and the straightforward application of bisection is no longer possible. A more general constraint formulation to model power flexibility amongst groups of antennas as well as to apply non-linear with respect to the power constraints can be found in \cite{Zheng2011a} and this generalization is part of future work.

\section{Multicast Multigroup Beamforming with Per Antenna Power Constraints}\label{sec: problem}
The focus is on the Max-Min Fair  problem with per antenna constraints which reads as
\begin{empheq}[box=\fbox]{align}
\mathcal{F:}\    \max_{\  t, \ \{\mathbf w_k \}_{k=1}^{G}}  &t& \notag\\
\mbox{subject to } & \frac{|\mathbf w_k^\dag \mathbf h_i|^2}{\sum_{l\neq k }^G |\mathbf w_l^\dag\mathbf h_i|^2+\sigma_i^2}\geq t, &\label{const: F SINR}\\
 \text{and to }\ \ \ \ & \left[\sum_{k=1}^G  \mathbf w_k\mathbf w_k^\dag  \right]_{nn}  \leq P_n, \label{const: F power}\\
\forall i \in\mathcal{G}_k, &k,l\in\{1\dots G\}, \forall n\in \{1\dots N_{t}\},\notag
 \end{empheq}
 where $\mathbf w_k\in \mathbb{C}^{N_t}$ and $t \in \mathbb{R}$.  Problem $ \mathcal{F,}$ with input $\mathbf p = [P_1, P_2\dots P_{N_t}]$, optimal value denoted as $t^*=\mathcal{F}(\mathbf p)$ and optimal point $\{\mathbf w_k^\mathcal{F}\ \}_{k=1}^{G}$, is non-convex due to  \eqref{const: F SINR}. The difference of the present formulation with respect to the Max-Min Fair problem with $\mathrm{SPC}$ presented in   \cite{Sidiropoulos2006,Karipidis2008} lies in \eqref{const: F power}, where there exist $N_t$ power constraints over each individual radiating element. This equation also differentiates the present formulation from the coordinated multicell multicasting Max-Min problem since {the per antenna constraint  is imposed on  the $n$-th   diagonal element of the summation of the correlation matrices of all precoders}. On the contrary, in \cite{Xiang2013}, the imposed per base station constraints are translated to one power constraint per each precoder.

The sum transmit power minimization problem under  $\mathrm{QoS}$ constraints\cite{Bengtsson2001}, is related to the Max-Min problem\cite{Wiesel-06} and this relation was generalized for the multicast multigroup case in \cite{Karipidis2008}. Hence by bisecting  the solution of the  $\mathrm{QoS}$ problem, a solution to the fairness problem can be derived. Nevertheless, two fundamental differences between the existing formulations and problem $\mathcal{F}$ complicate the solution. Firstly, the power constraints are not necessarily met with equality.
Rescaling the power of one precoding vector affects  the power inequality constraints of all antennas. Consequently, if only one constraint is satisfied with equality and some power budget is still left, a rescaling to approach the rest $N_t - 1$ constraints will over-satisfy the first constraint and render the problem infeasible.   Secondly, the absence of a related, solvable  problem prohibits the immediate application of bisection.

To the end of providing a framework for the per antenna constrained Max-Min Fair precoding  similar to \cite{Karipidis2008} and by elaborating on the insights of  \cite{Xiang2013},   a  novel per antenna power utilization minimization problem is defined as
\begin{empheq}[box=\fbox]{align}
\mathcal{Q:} \min_{\ r, \ \{\mathbf w_k \}_{k=1}^{G}}  &r& \notag\\
\mbox{subject to } & \frac{|\mathbf w_k^\dag \mathbf h_i|^2}{\sum_{l\neq k }^G | \mathbf w_l^\dag\mathbf h_i|^2 + \sigma^2_i}\geq t \label{const: Q SINR}\\
\text{and to}\ \ \ \ \ &\frac{1}{P_n} \left[\sum_{k=1}^G  \mathbf w_k\mathbf w_k^\dag \right]_{nn} \leq  r\\
\forall i \in\mathcal{G}_k,&k,l\in\{1\dots G\}, \forall n\in \{1\dots N_{t}\}.\notag
 \end{empheq}
 Problem $\mathcal{Q }$ receives as input a common $\mathrm{QoS}$ constraint to all users, namely $t,$ and the per antenna power constraint vector $\mathbf p = [P_1, P_2\dots P_{N_t}]$. Subsequently,  the maximum  power consumption out of all antennas is minimized and this solution can be denoted as $r^*=\mathcal{Q}(t, \mathbf p).$

 \textit{Claim 1}: Problems $\mathcal{F}\ $ and  $\mathcal{Q}\ $ are related as follows
\begin{align}
1= \mathcal{Q}\left(\mathcal{F}\left(\mathbf p\right),\mathbf p\right)\label{eq: equivalence 1}\\
t= \mathcal{F}\left(\mathcal{Q}\left(t, \mathbf p\right)\cdot\mathbf p\right)\label{eq: equivalence 2}
 \end{align}
 \textit{Proof: }The above claim will be proven by contradiction. Let $r^*=\mathcal{Q}(t, \mathbf p)$ denote the optimal value of $\mathcal{Q}$ with associated variable $ \{\mathbf w_k^Q \}_{k=1  }^{G}$. Assuming that the optimal value of $\mathcal{F}$ under constraints scaled by the solution of $\mathcal{Q}$ is different, i.e. $\hat t= \mathcal{F}\left(\mathcal{Q}\left(t, \mathbf p\right)\cdot\mathbf p\right)$ with $ \{\mathbf w_k^F \}_{k=1   }^{G}$, the following contradictions arise. In the case where $\hat t < t $, then the precoders   $ \{\mathbf w_k^Q \}_{k=1  }^{G}$  are feasible solutions to  $\mathcal{F}\left(\mathcal{Q}\left(t, \mathbf p\right)\cdot\mathbf p\right)$ which lead to a higher minimum $\mathrm{SNIR}$, thus contradicting the optimality of $\hat t$. Alternatively, if $\hat t > t $ then the solution set $ \{\mathbf w_k^F \}_{k=1   }^{G}$ can be scaled by a positive constant $ c = t /\hat t<1$. The new solution  $ \{c \mathbf w_k^F \}_{k=1   }^{G}$ respects the feasibility conditions of $\mathcal{Q}$ and provides a lower optimal value, i.e. $c\cdot r^*$, thus again contradicting the optimality hypothesis of $ \{\mathbf w_k^Q \}_{k=1  }^{G}$. The proof of  \eqref{eq: equivalence 1} follows an identical line of reasoning and is omitted for shortness. $\square$

Having defined an equivalent problem that admits a solution via the well know $\mathrm{SDR}$ technique, the original Max-Min Fair problem with $\mathrm{PAC}$ can be solved via a one dimensional bisection search over problem  $\mathcal{Q}$. More details on bisection are given in Sec. \ref{sec: bisec}.

%
%

\subsection{Semidefinite Relaxation }\label{sec: SDR}

 Problem  $\mathcal{Q}$ belongs to the general class of non-convex $\mathrm{QCQP}$s for which the $\mathrm{SDR}$ technique has proven a powerful and computationally efficient approximation technique. In the spirit of \cite{Bengtsson2001},  $\mathcal{Q}$  can be rewritten by using the change of variables $\mathbf X_i = \mathbf w_i \mathbf w_i^\dag $ and  introducing two additional constraints. Hence, the new variable is constrained to be symmetric positive semi-definite and unit-rank. However, the latter is a non-convex constraint.  The $\mathrm{SDR}$ method consists of dropping the unit-rank constraint and reducing the original problem into
\begin{empheq}[box=\fbox]{align}
\mathcal{Q}_r:\min_{r,\ \{\mathbf X_k \}_{k=1}^{G}}  &r& \notag\\
\mbox{subject to } & \frac{\mathrm{Tr}\left(\mathbf Q_i^\dag \mathbf X_k\right)}{\sum_{l\neq k }^G \mathrm{Tr}\left(\mathbf Q_l^\dag \mathbf X_k\right) +\sigma_i^2}\geq t \label{const: Q_r SINR}\\
 \text{and to} \ \ \ \ \ &\frac{1}{P_n} \left[\sum_{k=1}^G \mathbf X_{k}\right]_{nn} \leq  r   \label{const: Q_r Power}\\
 \text{and to}  \ \ \ \ \ &\ \mathbf X_k\succeq0,\label{const: Q_r SDF} \\
\forall i \in\mathcal{G}_k,&k,l\in\{1\dots G\}, \forall n\in \{1\dots N_{t}\},\notag
 \end{empheq}
\begin{empheq}[box=\fbox]{align}
\mathcal{Q}:\min_{r,\ \{\mathbf X_k \}_{k=1}^{G}}  &r& \notag\\
\mbox{subject to } & \frac{\mathrm{Tr}\left(\mathbf Q_i^\dag \mathbf X_k\right)}{\sum_{l\neq k }^G \mathrm{Tr}\left(\mathbf Q_l^\dag \mathbf X_k\right) +\sigma_i^2}\geq t \label{const: Q_r SINR}\\
 \text{and to} \ \ \ \ \ &\frac{1}{P_n} \left[\sum_{k=1}^G \mathbf X_{k}\right]_{nn} \leq  r   \label{const: Q_r Power}\\
 \text{and to}  \ \ \ \ \ &\ \mathbf X_k = \mathbf w_k \mathbf w_k^\dag \succeq0,  \label{const: Q_r SDF} \\
 \mbox{and to }\ \ \ \ &  \mathrm{rank}\left(\mathbf X_k^*\right) \neq 1
  \\\
\forall i \in\mathcal{G}_k,&k,l\in\{1\dots G\}, \forall n\in \{1\dots N_{t}\},\notag
 \end{empheq}

where $\mathbf Q_i = \mathbf h_i\mathbf h_i^\dag$.  $\mathcal{Q}_r$ is convex and can therefore be solved to an arbitrary accuracy  \cite{convex_book}.
In the same direction, the Max-Min Fair optimization can be also relaxed as
\begin{empheq}[box=\fbox]{align}
\mathcal{F}_r:\max_{t, \ \{\mathbf X_k \}_{k=1}^{G}}&t& \notag\\
\mbox{subject to } & \frac{\mathrm{Tr}\left(\mathbf Q_i^\dag \mathbf X_k\right)}{\sum_{l\neq k }^G \mathrm{Tr}\left(\mathbf Q_l^\dag \mathbf X_k\right)+ \sigma_i^2}\geq t \label{const: F_r SINR}\\
\mbox{and to }\ \ \ \ &\left[\sum_{k=1}^G  \mathbf X_k \right]_{nn} \leq  P_n\\
 \mbox{and to }\ \ \ \ &  \mathbf X_k \succeq0,
  \\\
\forall i \in\mathcal{G}_k,&k,l\in\{1\dots G\}, \forall n\in \{1\dots N_{t}\}.\notag
  & \end{empheq}
which, however, remains non-convex due to the constraint \eqref{const: F_r SINR};
an obstacle that will be overcome in the remaining by acknowledging that the relaxed problems are also related by \eqref{eq: equivalence 1} and \eqref{eq: equivalence 2}.

The problems described so far belong to the general class of multigroup multicasting problems. Mathematically,  the main difference is that the number of precoding vectors  is equal to the number of groups and thus less than the number of users. This fact, renders these problems NP-hard and the semidefinite relaxation cannot provide a globally optimal solution to the original problem. Therefore, the approximations described in the following section need to be employed.
\subsection{Gaussian Randomization} \label{sec: Gaussian randomization}
For specific  optimization problems, the $\mathrm{SDR}$ provides globally optimum solutions. This  implies that the relaxed solution $\mathbf X^{*}$ has a unit rank. The most prominent example of this case is the optimal downlink beamforming under independent data transmission to all users \cite{Bengtsson2001}.  Nevertheless, due to the NP-hardness of the multicast problem,   the relaxed problems do not necessarily yield unit rank matrices. Consequently, one can  apply a rank-1 approximation over $\mathbf X^{*}$ and use the principal eigenvalue and eigenvector as an approximate solution to the original problem. Other types of rank-1 approximations are also possible depending on the nature of the original problem.

Despite the effectiveness and intuitive simplicity of any rank-1 approximation, the solution with the highest provable accuracy for the multicast case is given by the Gaussian randomization method \cite{Luo2010}. Let  $\mathbf X^{*}$ be a symmetric positive semidefinite solution of the relaxed problem. Then, a candidate solution to the original problem can be generated as a Gaussian random variable with zero mean and covariance equal to $\mathbf X^{*}$, i.e.  $\hat{\mathbf{w}}_k \backsim\mathbb{C}\mathbb{N}(0, \mathbf X^{*}_k ).  $  Nonetheless, an intermediate step between generating a Gaussian instance with the statistics obtained from the relaxed solution and creating a feasible candidate instance of the original problem still remains, since the feasibility of the original problem is not yet guaranteed.  This step is described in the following section.

Finally,  after generating a predetermined number of candidate solutions, the one that yields the highest objective value of the original problem can be chosen. The accuracy of this approximate solution is measured by the distance of the approximate objective value and the optimal value of the relaxed problem. This accuracy increases with the predetermined number of Gaussian randomizations.
\subsection{Feasibility Power Control }\label{sec: power control}
 Despite the wide applicability of the Gaussian randomization method, one has to bare in mind that it is a problem dependent procedure. After generating a random instance of a Gaussian variable with statistics defined by the relaxed problem, an additional step comes in play to guarantee the feasibility of the original problem. In \cite{Sidiropoulos2006},  a simple power rescaling of the candidate     solutions which follows the Gaussian randomization is sufficient to guarantee feasibility. Nevertheless, baring in mind that in the multigroup case  an interference scenario is dealt with,  different than in \cite{Sidiropoulos2006}, a simple rescaling does not guarantee feasibility. Therefore, an additional  optimization step is proposed in \cite{Karipidis2008} that distributes the power amongst the candidate precoders whilst guaranteeing the feasibility of the original problem. In the same direction,  a novel power control problem with per antenna power constraints is  defined herein. Given a set of  Gaussian instances, denoted as $\{\mathbf{\hat w}_k\}_{k=1 }^G$, the \textit{Multigroup Multicast Per Antenna power Control} ($\mathrm{MMPAC}$) problem is defined as
\begin{empheq}[box=\fbox]{align}
\mathcal{S^{\mathcal{F}}:} \max_{t, \ \{  p_{k} \}_{k=1}^{G}}  &t& \notag\\
\mbox{subject to } & \frac{|\mathbf{ w}_k^\dag \mathbf{  h}_i|^2  p_{k}}{  \sum_{l\neq k }^G|\mathbf{  w}_l^\dag \mathbf{ h}_i |^2   p_{l} +\sigma_i^2}\geq t \label{const: S F SINR}\\
\mbox{and to } \ \ \ \ & \left[\sum_{k=1}^G  \mathbf{\hat  w}_k\mathbf{\hat  w}_k^\dag p_k\right]_{nn} \leq   P_{n} \\
\forall i \in\mathcal{G}_k,&k,l\in\{1\dots G\}, \forall n\in \{1\dots N_{t}\},\notag
 \end{empheq}
with $\{  p_{k} \}_{k=1}^{G}>0$. Problem $\mathcal{S^{\mathcal{F}}}$ receives as input the per antenna power constraints and returns the maximum worst $\mathrm{SNIR}$ $t^{*} = \mathcal{S}(\mathbf p)$ and is also non convex in the fashion of $ \mathcal{F}$.

A very important remark is clear in the formulation of the power control problem. The optimization variable $\mathbf p $ is of size $G$, i.e. equal to the number of groups, while the power constraints are equal to the number of antennas, $N_t$. In each constraint, all the optimization variables contribute. This fact prohibits the total exploitation of the available power at the transmitter. Once at least one of the $N_t$ constraints is satisfied with equality and remaining power budget, then the rest can not be scaled up since this would lead to at least one constraint exceeding the maximum allowable value.
\subsection{Bisection}\label{sec: bisec}
As already mentioned, a solution to non-convex problems can be obtained by  iteratively solving a related problem at the midpoint of an interval that includes the original optimal value. To be more precise,    let us consider problems $\mathcal{Q}_r$ and $\mathcal{F}_r$, which are related as given in \eqref{eq: equivalence 1} and \eqref{eq: equivalence 2}.
Since $t$ represents the $\mathrm{SNIR}$ and is actually a fraction of two positive quantities, it will always be positive or zero. Also, if the system was interference  free while all the users had the channel of the best user, then the maximum worst $\mathrm{SNIR}$ would be $\max_i\{P_{tot}\mathbf Q_i/\sigma_i \}.$ These bounds do not affect the optimal solution as long as they include it. Further analytic investigation of a tighter upper bound can reduce the bisection iterations but is not included in the present work for brevity. By defining the interval $[L, U]$ with the minimum and maximum values of $\mathrm{SNIR}$, the solution of $r^* = \mathcal{Q}_r\left ( \left(L+U\right)/2, \mathbf p\right)$  is obtained. If this solution is lower than $1$ then the lower bound of the interval is updated with this value. Otherwise the value us assigned to the upper bound of the interval. Bisection is iteratively continued until an the interval size is reduced to a pre-specified value $\epsilon$. After a finite number of iterations the optimal value of $\mathcal{F}_r$ is given as the resulting value for which $L \text{ and } U$ become almost identical. This procedure provides an accurate solution to the relaxed non-convex Max-Min Fair problem.
Actually, this value is the upper bound of the minimum $\mathrm{SNIR}$ and it can be used to evaluate how close to the optimal solution the approximate solutions generated by Gaussian randomization are.

 The precoder correlation matrices for which the optimal value of the relaxed problem is achieved are possible solutions to the original problem. In the case that these solutions are unit rank, then the original problem is globally solved by the non zero eigenvalue and the corresponding eigenvector of the relaxed solution. Nevertheless, when this is not the case,  the Gaussian randomization described in Sec. \ref{sec: Gaussian randomization} is employed. Hence a predetermined number of random Gaussian solutions are generated. Following this, for each and every solution $\{\mathbf{\hat w}_k\}_{k=1 }^G$, the power of the precoders needs to be controlled so that feasible candidate solutions can be obtained. Once the $\mathrm{MMPAC}$ is solved, the candidate precoder can be calculated as
\begin{align}\label{eq: cand solution}
 \{ \mathbf{w}^*_k \}_{k=1}^G = \sqrt{\{  p_{k} \}_{k=1}^{G}} \cdot \{\mathbf{\hat w}_k\}_{k=1 }^G.
\end{align}
The final complication of the solution process lies in the non-convexity of
 $\mathcal{S}^F$. Hence,  bisection needs to be employed\footnote{A possible reformulation as a $\mathrm{GP}$ is not considered herein for the sake of brevity.} again over its convex equivalent
\begin{empheq}[box=\fbox]{align}
\mathcal{S^{\mathcal{Q}}:} \min_{r, \ \{ p_k \}_{k=1}^{G}}  &r& \notag\\
\mbox{subject to } & \frac{|\mathbf{\hat  w}_k^\dag \mathbf h_i|^2 p_{k}}{\sum_{l\neq k }^G  |\mathbf{\hat  w}_l^\dag\mathbf h_i|^2 p_{l}+\sigma_i^2}\geq t \label{const: S Q SINR}\\
\mbox{and to } \ \ \ \  &\frac{1}{  P_{n}} \left[\sum_{k=1}^G  \mathbf{\hat  w}_k\mathbf{\hat  w}_k^\dag  p_k\right]_{nn} \leq  r \\
\forall i \in\mathcal{G}_k,&k,l\in\{1\dots G\}, \forall n\in \{1\dots N_{t}\}.\notag
 \end{empheq}
 It should be noted that in this case, the bisection interval can be further reduced and thus the efficiency of the algorithm greatly improved by exploiting the fact that the maximum optimal value of  $\mathcal{S}^F$  cannot be greater than the optimal value of  $\mathcal{F}_r$ which has already been calculated. Thus the bisection interval is constrained between zero and $t^*$ of the relaxed problem and the number of iterations is reduced.
\subsection{Complexity}
 The complexity of the $\mathrm{SDR}$ and Gaussian randomization technique has been  exhaustively discussed in  \cite{Luo2010} and the references therein. In brief, the complexity can be considered as follows. The interior point methods that solve the $\mathrm{SDR}$ problems require at most $\mathcal{O}(\sqrt{LN_t}\log(1/\epsilon)) $, where $\epsilon $ is the desired accuracy, while the arithmetic operations for each iteration are not more than $\mathcal{O}({L^3 N_t^6 + KLN_t^2})$. Modern solvers such as the CVX tool \cite{convex_book} which calls  numerical  solvers such as SeDuMi for $\mathrm{SD}$ programs, also exploit the specific structure of matrices and the actual running time is reduced.
Furthermore, the bisection technique typically runs for $N_{iter} = [\log_2\left(U-L\right)<\epsilon]  $. Finally, the Gaussian randomization can be executed for an arbitrary number of iterations of course with increasing accuracy. Typically, 100 randomizations are performed \cite{Karipidis2008,Sidiropoulos2006}.
The general complexity of the proposed approach is  increased due to the  $N_t$ constraints but still in a polynomial and thus computationally efficient manner.
\subsection{ Summary}
The overall procedure  to acquire an approximate solution to the original multigroup multicast NP-hard Max-Min Fair problem under $\mathrm{PAC}$ is  summarized in Alg. \ref{Alg: MMF PAC}.

\begin{algorithm}[h]
 \SetAlgoLined \hrule
 \KwIn{$ N_{rand}, \mathbf p, \mathbf Q_i, \sigma_i^2 \ \forall i \in\{1\dots G\} $ }
 \KwOut{ Solution and optimal values of $\mathcal{F}_r$
 and approximate solutions and optimal values of $\mathcal{F}.$ }\hrule
 \Begin{
 \textbf{\textit{\textit{\uline{Step 1:}}}} Solve $\mathcal{F}_r$ by bisecting  $\mathcal{Q}_r$, (see Sec. \ref{sec: bisec}).
Let  this solution be $\{\mathbf w_{k}^{opt}\}_{k=1}^G$ and the associated
 $\mathrm{SNIR}$ as $t_{opt}$. This solution will be the upper bound for any  solution of $\mathcal{F}$.\\
 \eIf{$\mathrm{rank}(\mathbf X_{k}^{opt} )= 1, \forall \ k \in\{1\dots G\} $}
{the output is the dominant eigenvector and the max eigenvalue $\lambda_{max}(\mathbf X^{opt})$.  }
{
\textbf{\textit{\uline{Step 2:}}}
Gaussian randomization: generate $N_{rand}$  precoding vectors
$\{ \mathbf{\hat{w}}_{k}\}_{k=1}^G$, (see Sec. \ref{sec: Gaussian randomization} ).
$t^{*}_{(0)}= 0$\;
\For{$i=1\dots N_{rand}$}{
\textbf{\textit{\uline{Step 3:}}}
Solve $\mathcal{S}^{\mathcal{F}}$ by bisecting the related  $\mathcal{S}^{\mathcal{Q}}$. The corresponding solution $\{\mathbf w_{k}^{can}\}_{k=1}^G$ reads as in \eqref{eq: cand solution} with associated optimum value $t_{(i)}^*$.\\
 \If {$t_{(i)}^* > t_{(i-1)}$ }{the current solution becomes the output\;}
}
}
}\hrule  \label{Alg: MMF PAC}
 \caption{ Max-Min-Fair Multigroup Multicasting under Per Antenna power Constraints.}
\end{algorithm}

\section{Numerical results} \label{sec: simulations}
The performance of Linear Multicast Multigroup beamforming under per antenna power constraints is presented herein. A system with $N_t = 5$ transmit antennas, $G = 2 $ groups and  $N_u = 4  $ users is assumed unless stated otherwise.   Rayleigh fading is considered, thus the channel instances are generated as Gaussian complex variable instances with unit variance and zero mean. For every channel instance, the   solutions of the Max-Min Fair $\mathrm{SPC}$  \cite{Karipidis2008} and the proposed $\mathrm{PAC}$ problems are evaluated using $N_{rand} = 50$ Gaussian randomizations.
 Noise variance is normalized to one for all receivers.

\begin{figure}[h]
\centering
\includegraphics[width=0.99\columnwidth]{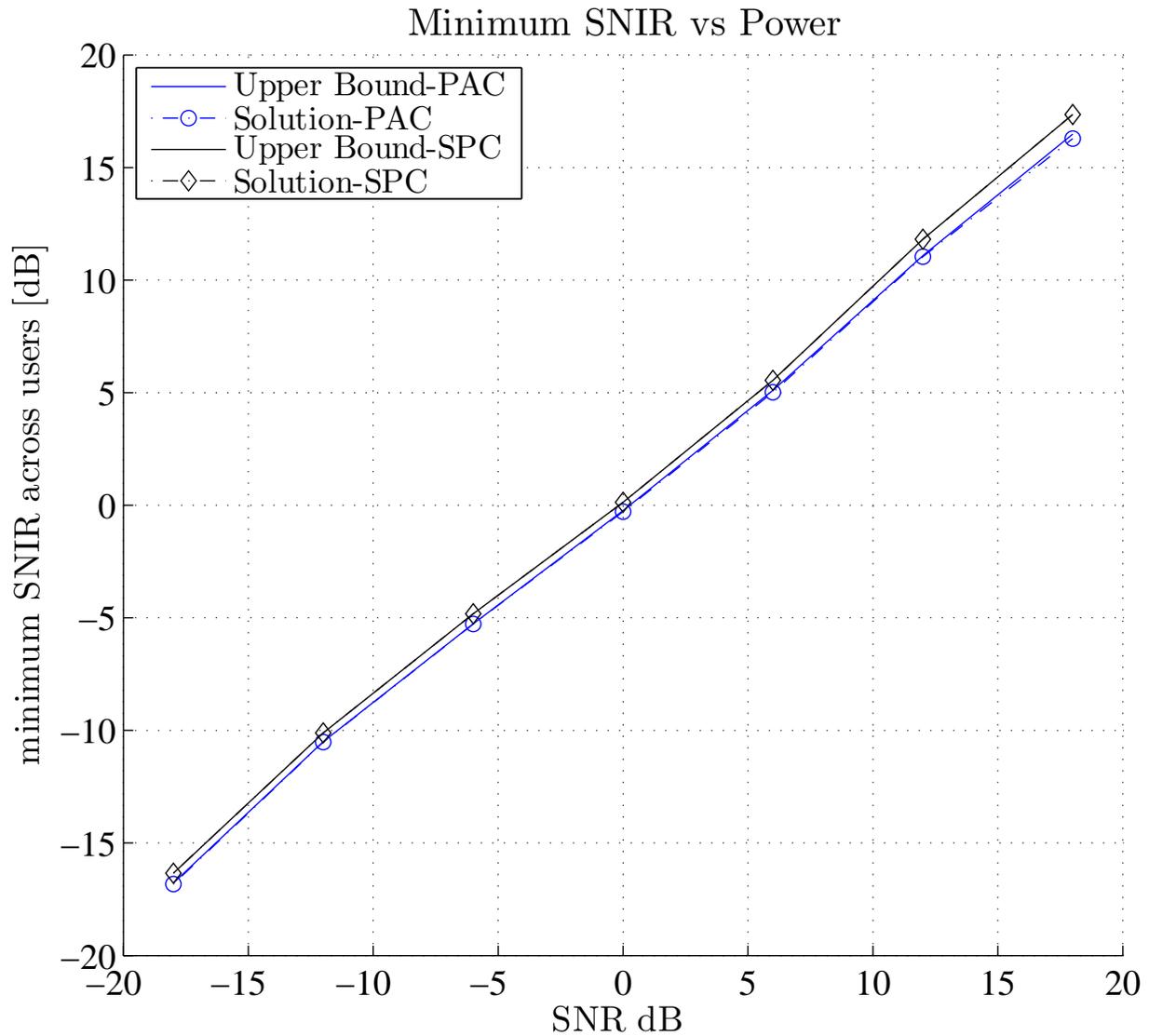}\\
\caption{Minimum $\mathrm{SNIR}$ with $\mathrm{SPC}$ and $\mathrm{PAC}$  versus increasing  total transmit power over the noise level ($\mathrm{SNR}$ [dB]).}\label{fig:  power}
\end{figure}

Firstly,  the accuracy of the approximate solution is numerically shown. The Max-Min Fair minimum $\mathrm{SNIR}$  under $\mathrm{SPC}$ and $\mathrm{PAC}$ constraints is plotted in Fig. \ref{fig:  power}
 with respect to the total transmit power to receive noise ratio $\mathrm{(SNR)}$ in dB. For fair comparison, the total power constraint $P_{tot}$~[Watts] is equally distributed amongst the transmit antennas when $\mathrm{PAC}$ is considered, hence each antenna can radiate at most $P_{tot}/N_t $~[Watts].  The accuracy of the approximate solutions for both problems  is clear across a wide range of on board power. Nevertheless, the accuracy due to the per antenna constraints is insignificantly reduced. This is justified by the fact that a Gaussian randomization instance is less likely to approach the optimal point when the number of constraints is increased. It is reminded that both problems are solved under the same number of Gaussian randomizations ($ N_{ rand} = 50 $).

\begin{figure}[h]
\centering
\includegraphics[width=0.99\columnwidth]{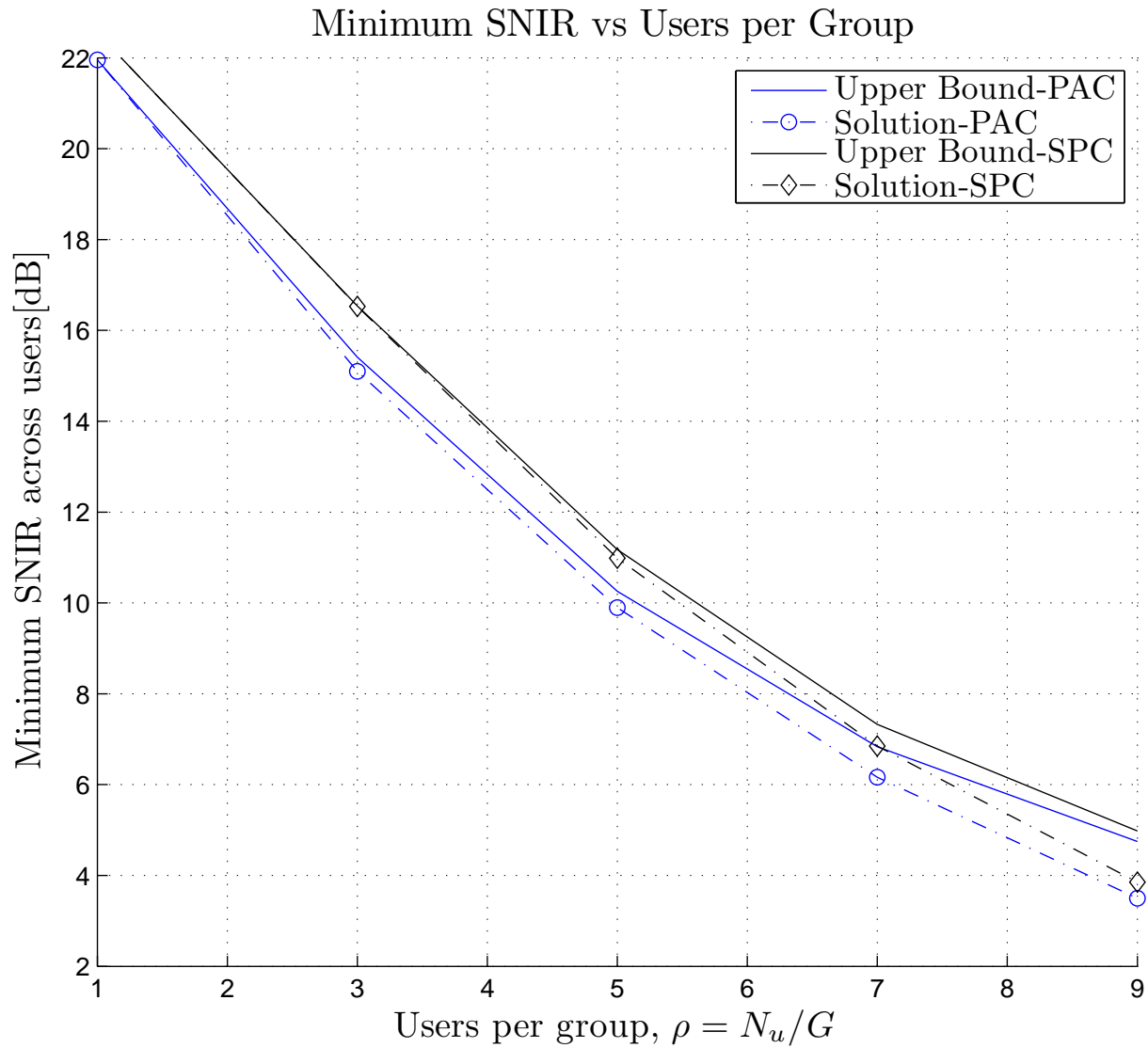}\\
\caption{Minimum $\mathrm{SNIR}$ with $\mathrm{SPC}$ and $\mathrm{PAC}$  versus increasing  ratio of users per group $ \rho = N_u / G  $.   }\label{fig:  users}
\end{figure}

A significant discussion over the $\mathrm{SDR}$ techniques in multicast applications is the scaling of the approximate solution to the NP-hard problem versus an increasing number of receivers per multicast. In the extreme case of one user per group, it was proven in\cite{Bengtsson2001} that the relaxation provides an optimal solution. Thus the solution is no longer approximate but exact. However, the increasing number of users per group degrades the solution, as depicted in Fig. \ref{fig:  users} for both problems. It is especially noticed that the $\mathrm{PAC}$ system suffers equivalently with respect to the $\mathrm{SPC}$ of \cite{Karipidis2008} as the number of users per multicast group increases.

\section{Conclusions} \label{sec: conclusions}
The problem of optimizing the linear precoding design under per antenna power constraints, when common data is addressed to multiple co-channel groups is tackled in the present work. A novel framework  to find an approximate solution to the NP-hard multigroup multicast problem with $\mathrm{PAC}$  is proposed. Under this framework, the  linear precoders for the multi-antenna systems with limited power in each transmitting element, can be accurately approximated in polynomial time. Consequently, an important practical constraint towards the implementation of physical layer multigroup multicasting is alleviated.

\bibliographystyle{IEEEtran}
\bibliography{refs/IEEEabrv,refs/conferences,refs/journals,refs/books,refs/references,refs/csi,refs/thesis}
\end{document}